\documentclass[reprint,nofootinbib,superscriptaddress,amsmath,amssymb,aps]{revtex4-1}
\usepackage{amsmath,amsfonts,amssymb,amsthm,mathrsfs,latexsym,epsfig,cancel,enumerate,times}  
\usepackage{xcolor}
\definecolor{navy}{RGB}{0,0,150}
\usepackage[colorlinks, linkcolor=cyan, citecolor=navy, urlcolor=navy, plainpages=false, pdfstartview=FitH]{hyperref}
\usepackage{appendix}
\allowdisplaybreaks
\def\be{\begin{equation}}
\def\ee{\end{equation}}
\def\ba{\begin{eqnarray}}
\def\ea{\end{eqnarray}}
\def\nn{\nonumber}

\usepackage{color} 

\newcommand{\abs}[1]{{\left|{#1}\right|}} 
\newcommand{\inner}[2]{{\langle {#1}\vert {#2} \rangle}} 
\newcommand{\ket}[1]{\vert{#1}\rangle} 
\newcommand{\bras}[1]{({#1}\vert} 
\newcommand{\bra}[1]{\langle{#1}\vert} 

\newcommand{\ints}{{\int_\Sigma}} 

\newcommand{\hil}{\mathcal{H}} 

\begin{document}


\title{ Realization of Dirac quantization in loop quantum gravity}
\author{Xiangdong Zhang}\email{scxdzhang@scut.edu.cn}
\affiliation{Department of Physics, South China University of
Technology, Guangzhou 510641, China}
\author{Yongge Ma}
\thanks{Corresponding author}
\email{ mayg@bnu.edu.cn}
\affiliation{Department of Physics, Beijing Normal University, Beijing 100875, China}

\begin{abstract}
The system of gravity coupled to the non-rotational dust field is studied at both classical and quantum levels. The scalar constraint of the system  can be written in the form of a true physical Hamiltonian with respect to the dust time. In the framework of loop quantum gravity, the scalar constraint is promoted to a well-defined operator in a suitable Hilbert space of the coupled system, such that the physical Hamiltonian becomes a symmetric operator. By the deparametrized form, a general expression of the solutions to the quantum scalar constraint is obtained, and the observables on the space of solutions can be constructed. Moreover, the Dirac quantization procedure can be fully carried out in loop quantum gravity by this system.
\end{abstract}


\maketitle

It is well known that due to the singularities of the big bang and the
black hole interior, classical general relativity (GR) is no longer valid at the regime where the spacetime curvature becomes divergent. Finding a consistent theory of quantum gravity serves as one of the main driving forces in theoretical physics in the past decades \cite{QG1,QG2,QG3}, and various approaches have been
pursued, including string/M-Theory \cite{string1,string2,string3,string4} and loop quantum
gravity(LQG) \cite{Ro04,Th07,As04,Ma07}. LQG is notable by its background
independent feature. This background
independent quantization method has been successfully generalized to a few modified theories of gravity \cite{Zh11,Zh11b,Ma18,Zhang20}.

The notion of time plays
an important role in any quantum gravity theories, since in diffeomorphism-invariant theories one only has the Hamiltonian constraint rather than a true Hamiltonian which represents the evolution of the system \cite{Isham94}. To overcome the time problem in quantum gravity \cite{Isham94}, one usually takes the viewpoint of relational evolution and employ the deparametrization technique \cite{Lewandowski10,Lewandowski15,RS94,Kuchar91}. This
allows one to map the totally constrainted theory of canonical GR into a theory
with a true nonvanishing Hamiltonian with respect to some chosen dynamical (emergent) time variable. The deparametrization formalism were realized to a certain extent in a few systems, including massless scalar field \cite{RS94,Lewandowski10,Lewandowski15} and dust fields \cite{Kuchar91,Husain15,Thiemann15}. The combination of LQG with the deparametrization
framework makes it possible to solve the quantum Hamiltonian constraint. In the literature, there exist two different strategies to solve the coupled Hamiltonian constraint. The first strategy is to impose some gauge conditions or solve some constraint (usually the diffeomophism constraint) at the classical level to simplify the model, and then to deparametrize the Hamiltonian constraint and quantize the model \cite{Husain15,Thiemann10}. In \cite{Husain15}, a particular time gauge $t=T$ is adopted, where $T$ is the configuration variable of the non-rotational dust while $t$ represents the time of the system. This means that the time reparametrization is no longer a gauge symmetry \cite{Pawlowski12,Lewandowski17}. Also,
the authors in \cite{Lewandowski10,Thiemann15} consider another possibility of using the diffeomophism constraint to re-express the Hamiltonian constraint and then quantizing the deparametrized Hamiltonian. This treatment usually requires to solve the diffeomophism constraint at the classical level or restrict all discussions on the diffeomorphism-invariant quantum states. However, how to express the gravitational diffeomorphism constraint as an operator is still unclear in this treatment.

In the second strategy proposed in \cite{Lewandowski15}, one quantizes the coupled system of gravity and a massless scalar field in the usual way of LQG, and then deparametrizes the system after quantization and tries to obtain solutions of the quantum constraints. The merit of this strategy is that the full Hamiltonian constraint  is realized at quantum level, and it is possible to define a true Hamiltonian operator to represent evolution and to find the physical solutions. However, in this model, the commutator of two Hamiltonian operators does not vanish and hence no nontrival solutions could be obtained. In the present letter, we will extend this favorable strategy to the coupled system of gravity and the non-rotational dust field which was regarded  as a realistic matter field to deparametrize GR \cite{Pawlowski12,Thiemann15,Lewandowski17,Husain15,Thiemann10}. Our purpose is to quantize the coupled system without imposing any gauge fixing or solving any constraint before quantization. We will show that the scalar constraint can be promoted to a well-defined operator in a suitable Hilbert space of the coupled system, such that the physical Hamiltonian becomes a symmetric operator. By the deparametrized form, a general expression of the solutions to the quantum scalar constraint is obtained for the first time, and the observables on the space of solutions can be constructed.

The action for the non-rotational dust coupled to gravity reads \cite{Kuchar95,Pawlowski12,Thiemann15,Lewandowski17,Husain15,Thiemann10}
 \ba
S&=&\frac{1}{2}\int d^4x\sqrt{-g}\left[\frac{1}{\kappa}R+M\left( g^{ab}(\partial_{a}T)\partial_{b}T+1\right)\right],
\ea where $g$ denotes the determinant of the spacetime metric $g_{ab}$, $M$ is the rest mass density of the dust field and $\kappa=8\pi G$ with $G$ being the Newton's gravitational constant. In the
Hamiltonian formalism, the diffeomorphism and Hamiltonian constraints of the coupled system read respectively
\begin{align}
C_a(x)&=C^{gr}_{a}(x)+\pi(x)T_{,a}(x)=0,\label{dcgd}\\
C^{tot}&=C^{gr}(x)+\frac{1}{2}\left(\frac{\pi^2}{M\sqrt{q}}+M\sqrt{q}\left(1+q^{ab}T_{,a}T_{,b}\right)\right)\nn\\
&=0,\label{HamiltonGD}
\end{align}
where $T_{,a}\equiv\partial_{a}T(x)$, $q$ denotes the determinant of the spatial metric $q_{ab}$, $C^{gr}_{a}(x)$ and $C^{gr}(x)$ are the gravitational diffeomorphism constraint and Hamiltonian constraint of GR respectively, $\pi(x)$ is the conjugate momentum of $T(x)$ satisfying \cite{Pawlowski12,Lewandowski17,Husain15} \ba
\pi(x)=\pm M\sqrt{q}\sqrt{1+q^{ab}T_{,a}(x)T_{,b}(x)}\label{pi}.
 \ea Following the same convention in \cite{Thiemann15,Husain15} and substituting Eq. \eqref{pi} into \eqref{HamiltonGD}, one obtains \ba
C^{tot}&=\abs{C^{gr}(x)}+\pi(x)\sqrt{1+q^{ab}T_{,a}(x)T_{,b}(x)}=0.\label{Cbefore}
\ea Since the variable $T$ represents the proper time of the dust particles, one has $1+q^{ab}T_{,a}T_{,b}\neq 0$. Hence the constraint \eqref{Cbefore} is equivalent to \cite{Thiemann15}\ba
C^{tot}=\pi(x)+\frac{\abs{C^{gr}(x)}}{\sqrt{1+q^{ab}T_{,a}T_{,b}}}=0. \label{CHC}
\ea While the term $\mathscr{T}\equiv\frac{1}{\sqrt{1+q^{ab}T_{,a}T_{,b}}}$ looks an obstacle for the quantization of \eqref{CHC}, our key observation is that it can be re-expressed in a form suitable for the loop quantization.

In the connection-dynamical formalism of GR, the basic canonical pair for gravity is the $su(2)$-valued connection $A^i_a$ and the densitizd triad $E^a_i$ with basic Poisson bracket \cite{Th07}\ba
\{A^i_a(x),E^b_j(y)\}=\kappa\beta\delta^b_a\delta^i_j\delta(x,y),
\ea where $\beta$ is the Barbero-Immirzi parameter. Based on the connection formalism of GR, the kinematical structure of LQG has been constructed rigorously \cite{As04,Ma07}. However, the quantum dynamics of LQG encoded in the Hamiltonian constraint remains an open issue. The Hamiltonian constraint operators proposed in \cite{Lewandowski15,Lewandowski15b} do not generate new vertices on the graphs of the cylindrical functions and hence are symmetric in the Hilbert space consisting of the states that are diffeomorphism invariant up to the vertices of their graphs. Another symmetric Hamiltonian constraint operator was also been proposed in \cite{Ma15}, which does generate new vertices and is well defined in the Hilbert space of the state that are diffeomorphism invariant up to the non-planar vertices with valence
higher than three. Thus there are consistent ways to define the Hamiltonian constraint operator corresponding to $C^{gr}$ in Eq.\eqref{CHC}. To overcome remaining obstacle of the term $\mathscr{T}$, we notice the following classical identity\begin{widetext}\ba
\mathscr{T}&=&\frac{\sqrt{q}}{\sqrt{q+E_i^aE_i^bT_{,a}T_{,b}}}=\frac{2}{\kappa\beta\sqrt{q}}E_i^a
\{A_a^i,S[1]\}-\frac{2S}{\sqrt{q}},\label{Thiemanntrick}
\ea
\end{widetext}where $q=\frac{1}{3!}\abs{\varepsilon_{abc}\varepsilon^{ijk}E_i^aE_j^bE_k^c}$ and $S[f]\equiv\int d^3xf(x)S(x):=\int d^3x f(x)\sqrt{q(x)+E_i^aE_i^bT_{,a}T_{,b}}$. Therefore, the smeared version of Hamiltonian constraint \eqref{CHC} reads
\ba
C(N)&=&\ints d^3x N(\pi(x)+h(x))=0\label{HamiltonGD1}
\ea where
\ba
h(x)&\equiv&\abs{C^{gr}(x)}\left(\frac{2}{\kappa\beta\sqrt{q}}E_i^a\{A_a^i,S[1]\}-\frac{2S(x)}{\sqrt{q}}\right)\nn\\
&=:&h_1(x)+h_2(x).\label{h1+h2}
\ea Eq.\eqref{HamiltonGD1} implies that one can define a physical Hamiltonian $h(x)$ which generates the evolution of the system with respect to the dynamical dust "time" $T$.

To quantize the non-rotational dust, we will employ the polymer quantization such that the "exponent" of the integrated momentum of $T$ is well represented \cite{Lewandowski15}, which is suitable to construct an operator corresponding to \eqref{Thiemanntrick}. Thus the basic variables are quantized as \cite{Lewandowski15}
\ba
\hat{T}(x)\ket{\phi}=\phi(x)\ket{\phi},\\
\exp{(-\frac{i}{\hbar}\int d^3x p(x)\hat{\pi}(x))}\ket{\phi}=\ket{\phi+p},
\ea where $\ket{\phi}$ represents a normalized basis in the Hilbert space $\hil_T$ of the dust field, such that
\ba
\phi[\pi]:=\inner{\pi}{\phi}=\exp{\left(-\frac{i}{\hbar}\int d^3x \phi(x)\pi(x)\right)}.
\ea The diffeomorphisms $\varphi$ of $\Sigma$ act unitarily in $\hil_T$ by
\ba
\hat{U}_\varphi\phi[\pi]=\phi[\varphi^*\pi]=(\phi\circ\varphi^{-1})[\pi].
\ea Moreover, since $\ket{\phi}$ are eigenstates of the self-adjoint operator $\hat{T}(x)$, the operator $\widehat{T,_{a}}(x)$ corresponding to $\partial_aT(x)$ can be defined as\ba
\widehat{T,_{a}}(x)\ket{\phi}=\left(\partial_a\phi(x)\right)\ket{\phi}.
\ea
The kinematical Hilbert space is a direct product $\hil_{kin}=\hil_{T}\otimes\hil_{gr}$ of the dust field part $\hil_{T}$ and the geometry part $\hil_{gr}$ with the orthonormal basis $\ket{\phi}\otimes \ket{\gamma,j,i}$, where $\gamma$ denotes a given finite graph, $j$ labels the $SU(2)$ representations associated to the edges of $\gamma$, and $i$ labels the intertwiners assigned to the vertices linking the edges \cite{Th07,Lewandowski15}. We note that the Gaussian constraint can be easily solved by the gauge invariant spin-network states as in LQG, so that the gauge invariant kinematical Hilbert space $\hil^{G}$ for the coupled system can be obtained. The orthonormal basis in $\hil^{G}$ will be still denoted as $\ket{\phi}\otimes \ket{\gamma,j,i}$. To define a quantum scalar constraint operator corresponding to \eqref{HamiltonGD1},
as a first step, we will try to define an operator $\hat{h}(x)$ in $\hil^{G}$. By the point-splitting method, the smeared term of $h_2$ in \eqref{h1+h2} can be regularized as\begin{widetext}\ba
\int d^3xN(x)h_2(x)&=&-\lim_{\varepsilon\rightarrow 0}\ints d^3x \frac{2}{V_{U_x^\varepsilon}}N(x)\abs{C^{gr}(x)}\ints d^3y \chi_\varepsilon(x-y) S(y),\label{integralh2}
\ea
\end{widetext}where $\chi_\varepsilon(x-y)$ is the characteristic function with width $\varepsilon$, $V_{U_x^\varepsilon}$ denotes the volume of an arbitrary
neighborhood $U_x^\varepsilon$ containing the point $x$ with the size $\varepsilon$. To deal with the second integral in the right hand side of Eq. \eqref{integralh2},
we first notice that the classical expression $\int d^3x f(x)\sqrt{E^a_iE^b_iT_{,a}T_{,b}}$ can be defined as an operator in $\hil^G$ as \cite{Ma05,Lewandowski15}
\begin{widetext}
\ba
\int d^3x f(x)\sqrt{\widehat{E^a_iE^b_iT_{,a}T}_{,b}}\cdot \ket{\phi}\otimes \ket{\gamma,j,i}
&=&8\pi \gamma \ell^2_p\left(\sum_I\sqrt{j_I(j_I+1)}\sum_{s=1}^{m_I}\int_{e_I^{(s)}}fd\phi\right)\cdot \ket{\phi}\otimes \ket{\gamma,j,i},\label{MaLingO}
\ea
\end{widetext}
where $I$ ranges the labels of the edges
of $\gamma$, and $e^{(1)}_I,...e^{(k)}_I , ..., e^{(mI)}_I$ are segments of $e_I$ such
that $\phi|_{e^{(k)}_I}$
is monotonic, and they are oriented in such
a way that the scalar field $\phi$ is growing along each of them. Then, to regularize this integral denoted by $S[\chi_\varepsilon]$, we introduce a family of
partitions of $\Sigma$, which is parameterized by some scale $\delta$ such that \ba
\Sigma=\bigcup_r\Sigma_r^\delta.
\ea Then in the limit of $\delta\rightarrow 0$, the integral of $S[\chi_{\epsilon}]$ can be taken into the two terms respectively  before taking the quare-root \cite{Lewandowski15}.
Moreover, one can suitably choose the intertwiners $i$ such that the basis $\ket{\phi}\otimes \ket{\gamma,j,i}$ consists of the eigenstates of the volume operator as well as the operator in Eq.\eqref{MaLingO}. Then one obtains
\begin{widetext}
\ba
&&\int_\Sigma d^3y \chi(x-y)\sqrt{\widehat{q}+\widehat{E^a_iE^b_iT_{,a}T_{,b}}}\cdot\ket{\phi}\otimes \ket{\gamma,j,i}\nn\\
&&=\left[8\pi \beta \ell^2_p\left(\sum_I\sqrt{j_I(j_I+1)}\sum_{s=1}^{m_I}\int_{e_I^{(s)}}\chi_\varepsilon(x-y)d\phi\right)+\sum_{\alpha'} V_{v_{\alpha'}}\chi_\varepsilon(x-v_{\alpha'})\right]\ket{\phi}\otimes \ket{\gamma,j,i},
\ea
\end{widetext} where $V_{v_{\alpha'}}$ denotes the eigenvalue of the volume operator at the vertex $v_{\alpha'}$ of the graph $\gamma$. Since the regulated gravitational Hamiltonian constraint operator $\hat{C}^{gr}_{\delta}$ acts only on the vertices, Eq. \eqref{integralh2} can be promoted as the following operator
\begin{widetext}
\ba
&&\int_\Sigma d^3x N(x)\hat{h}^{\delta}_2(x)\cdot\ket{\phi}\otimes \ket{\gamma,j,i}\nn\\
&&=-2\lim_{\varepsilon\rightarrow0}\sum_\alpha \left(N(v_\alpha)\hat{V}^{-1}_{U^\varepsilon_{v_\alpha}}\abs{\hat{C}^{gr}_\delta(v_{\alpha})}\right)\left[8\pi \beta \ell^2_p\left(\sum_I\sqrt{j_I(j_I+1)}\sum_{s=1}^{m_I}\int_{e_I^{(s)}}\chi_\varepsilon(v_\alpha-y)d\phi\right)+ V_{v_\alpha}\right]\ket{\phi}\otimes \ket{\gamma,j,i}\nn\\
&&=-2\sum_\alpha \left(N(v_\alpha)\hat{V}^{-1}_{v_\alpha}\abs{\hat{C}^{gr}_\delta(v_{\alpha})} V_{v_\alpha}\right)\ket{\phi}\otimes \ket{\gamma,j,i},\label{finalactionh2}
\ea
\end{widetext} where $\hat{V}^{-1}_{v_\alpha}$ denotes the inverse volume operator at the vertex $v_{\alpha}$ \cite{Ma15,YangMa16}. Note that, in order to have a well-defined adjoint operator $\hat{C}^{gr}_{\delta}{}^{\dagger}$, we have used the freedom of choosing the spin representations attached to each new added loop by $\hat{C}^{gr}_{\delta}$ to ensure that the valence of any vertex of the spin network state would not be changed by its action \cite{Ma15,ZhangLewandowskiMa18}, and then $\abs{\hat{C}^{gr}_\delta(v_{\alpha})}$ should be understood as $\abs{\hat{C}^{gr}_\delta(v_{\alpha})}=\sqrt{\hat{C}^{gr}_\delta(v_{\alpha})\hat{C}^{gr}_\delta{}^{\dagger}(v_{\alpha})}$ \cite{Sahlmann15}. Note also that the last step in Eq. \eqref{finalactionh2} is taken because of $\lim_{\varepsilon\rightarrow0}\int_{e_I^{(s)}}\chi_\varepsilon(v_\alpha-y)d\phi=0$. It is surprising that the dust part dropped out of the final action of \eqref{finalactionh2} due to its coupling to the gravitational part.
Now we consider the quantization of the term $h_1$ in Eq.\eqref{HamiltonGD1}. The densitized triad can be expressed as $E^a_i=\frac12\varepsilon_{ijk}\varepsilon^{abc}e^j_be^k_c$ with
$e^j_b=\frac{2}{\beta}\{A^j_b(x),V(x)\}$, and by point-splitting the smeared term of $h_1$ can be regularized as
\begin{widetext}\ba
\ints d^3x N(x)h_1(x)
&=&\frac{2}{\kappa\beta}\lim_{\varepsilon\rightarrow 0}\ints d^3y \frac{E^a_i\{A_a^i,S[1]\}(y)}{V_{U_y^\varepsilon}}\chi_\varepsilon(x-y)\ints d^3x N(x)\abs{C^{gr}(x)}.\label{pointsplith1}
\ea
\end{widetext}Again, due to its coupling to the gravitational Hamiltonian $C^{gr}$, the dust part in the expression of $S[1]$ will drop out of the final action of the operator. Thus, the quantum analogy of \eqref{pointsplith1} acts on the basis states as
\begin{widetext}
\ba
&&\int_\Sigma d^3x N(x)\hat{h}^{\delta}_1(x)\cdot\ket{\phi}\otimes \ket{\gamma,j,i}\nn\\
&=&\frac{2}{\kappa\beta}\sum_{\alpha} \frac{8N(v_\alpha)}{\beta^2(i\hbar)^2}\varepsilon_{ijk}\varepsilon^{abc}\hat{h}^\delta_{s_j}[\hat{h}^{\delta-1}_{s_j},\sqrt{\hat{V}_{v_\alpha}}]
\hat{h}^\delta_{s_k}[\hat{h}^{\delta-1}_{s_k},\sqrt{\hat{V}_{v_\alpha}}]\hat{h}^\delta_{s_i}[\hat{h}_{s_i}^{\delta-1},\hat{V}_{v_\alpha}]
\abs{\hat{C}^{gr}_\delta(v_\alpha)}\cdot\ket{\phi}\otimes \ket{\gamma,j,i},\label{finalactionh1}
\ea
\end{widetext} where $\hat{h}^\delta_{s_j}$ is the holonomy along the segment $s_j$ starting at $v_\alpha$ with parameter length $\delta$.
Now the physical Hamiltonian $h(N)\equiv\ints d^3x N(x)h(x)$ has been promoted as an operator $\hat{h}^{\delta}(N)$ in $\hil^G$. The subtleties here are how to take the limit $\delta\rightarrow0$ and whether the final Hamiltonian operator could be symmetric. It turns out that the limit can be taken in the gravitational Hilbert space $\hil^{g}_{np4}$ of the spin-network states $\bras{\gamma,j,i}$ that are diffeomorphism invariant up to the non-planar vertices with valence higher than three \cite{Ma15}. Given a graph $\gamma$ underling a gauge invariant state $\ket{\gamma,j,i}\in\hil_{gr}$, one denotes its non-planar vertices with valence higher than 3 by $V_{np4}(\gamma)$, the group of all diffeomorphisms preserving $V_{np4}(\gamma)$ by Diff$(\Sigma)_{V_{np4}(\gamma)}$, and the diffeomorphisms acting trivially on $\gamma$ by TDiff$(\Sigma)_\gamma$. Then the states $\bras{\gamma,j,i}$ are defined as\ba
\bras{\gamma,j,i}\equiv \eta(|\gamma, j, i>):=\frac{1}{N_{\gamma}}\sum_{{\tiny{\varphi}}}\hat{U}_{\varphi}\cdot\ket{\gamma,j,i},
\ea where $\tiny{\varphi\in Diff(\Sigma)_{V_{np4}(\gamma)}/TDiff(\Sigma)_\gamma}$, $\hat{U}_{\varphi}$ is the unitary representation of $\varphi$ in $\hil_{gr}$, and $N_\gamma$ is a normalization factor. Since the states $\bras{\gamma,j,i}$ belong to the dual space $\mathscr{D}^*_g$ of a dense subset of $\hil_{gr}$, the inner product between the states in $\hil^g_{np4}$ is defined by $\inner{\eta(f)}{\eta(g)}:=\eta(f)(g)$. Thus, we could naturally define a symmetric physical Hamiltonian operator $\hat{h}^{sym}(N)=\frac12(\hat{h}(N)+\hat{h}^{\dag}(N))$ therein. Now we consider the quantization of the term $\int_\Sigma d^3x N(x)\pi(x)$ in Eq.\eqref{HamiltonGD1}. Although one can not define an operator corresponding to this term in $\hil_T$, such an operator does exist in the dual space $\mathscr{D}^*_T$ of a dense subset of $\hil_T$ with state $\bras{\varphi}$. The action of this operator reads \cite{Lewandowski15}
\begin{widetext}
\ba
\left[\left(\int d^3x\hat{\pi}(x)N(x)\right)^*\bras{\varphi}\right]\ket{\phi'}&&:=i\frac{d}{d\epsilon}\left[\exp\left(-i\epsilon\int d^3x\hat{\pi}(x)N(x)\right)^*\right]\sum_{\phi}\varphi[\phi]\bra{\phi}\ket{\phi'}\nn\\
&&=i\frac{d}{d\epsilon}\varphi[\phi'+\epsilon N]=:i\bra{\delta_N\varphi}\ket{\phi'}.
\ea
\end{widetext}
Hence, one has \ba
\hat{\pi}[N]\cdot\bras{\varphi}\equiv\left(\int d^3x\hat{\pi}(x)N(x)\right)^*\bras{\varphi}=i\bras{\delta_N\varphi}.\label{actionofpi}
\ea To be compatible with the gravitational Hilbert space $\hil^g_{np4}$, we consider only the states in $\mathscr{D}^*_T$ with the following form with respect to certain graphs $\gamma$,
\ba
\bras{\Psi,V_{np4}(\gamma)}=\sum_{\phi}\Psi(\phi(v_1),...,\phi(v_m))\bra{\phi},
\ea  where the function $\Psi$ depends only on values of $\phi$ taking on the non-planar vertices of $\gamma$ with valence higher than 3. It is obvious that $\bras{\Psi,V_{np4}(\gamma)}$ are invariant under the action of Diff$(\Sigma)_{V_{np4}(\gamma)}$. A natural inner produce can be defined for these states by \begin{widetext}
\ba
\bras{\Psi,V_{np4}(\gamma)}\Psi',V_{np4}(\gamma')):=\sum_{\phi,\phi'}\bar{\Psi}(\phi(v_1),...,\phi(v_m))\Psi'(\phi'(v_1),...,\phi'(v_m))
\delta_{\gamma,\gamma'}\inner{\phi}{\phi'}.\label{innerproduct}
\ea
\end{widetext}It turns out that the operator in \eqref{actionofpi} keeps the space of the states $\bras{\Psi,V_{np4}(\gamma)}$ invariant and hence is well defined in the Hilbert space $\hil^T_{np4}$ as the completion of $\mathscr{D}^*_T$ with respect to the inner product \eqref{innerproduct}. Thus the whole scalar constraint \eqref{HamiltonGD1} has been promoted as a well defined operator in $\hil^T_{np4}\otimes\hil^g_{np4}$.

We denote the states in $\hil^T_{np4}\otimes\hil^g_{np4}$ by $\bras{\Psi,\gamma,j,i}\equiv\bras{\Psi,V_{np4}(\gamma)}\otimes\bras{\gamma,j,i}$ and consider how to obtain the solutions to the quantum constraint {\small\ba
\left(\hat{\pi}[N]+\hat{h}^{sym}(N)\right)\cdot\bras{\Psi,\gamma,j,i}=0. \label{quantumconstraint}
\ea} Taking account of the fact that $\hat{h}^{sym}(N)$ commutates with $\hat{\pi}[N]$ and the following commutator \ba
\left[\hat{\pi}[N(x)],\hat{T}(y)^*\right]=iN(y),
\ea where $\hat{T}(y)^*$ is the dual of $\hat{T}(y)$ in $\hil^T_{np4}$, we have the following identity,
\begin{widetext}
\ba
\left(\hat{\pi}[N]+\hat{h}^{sym}(N)\right)\bras{\Psi,\gamma,j,i}=e^{i\int d^3x \hat{T}^*\hat{h}^{sym}}\hat{\pi}[N]e^{-i\int d^3x \hat{T}^*\hat{h}^{sym}}\bras{\Psi,\gamma,j,i}.\label{constrainteq}
\ea
\end{widetext} Therefore, the general solutions to the constraint equation \eqref{quantumconstraint} can be written as
\ba
\bras{\Psi,\gamma,j,i}=e^{i\int d^3x \hat{T}^*\hat{h}^{sym}}\bras{\Psi_0,\gamma,j,i},\label{physicalsolution}
\ea where the functional $\bras{\Psi_0,V_{np4}(\gamma)}$ satisfies
\ba
\bras{\delta_N\Psi_0,V_{np4}(\gamma)}=0.
\ea Following the ideas of the relational framework for observables \cite{Lewandowski10}, we can construct a large family of observables. Let $\hat{L}$ be a linear operator in $\hil_{gr}$, which is invariant with respect to the action of Diff($\Sigma$)$_{V_{np4}(\gamma)}$. Then its dual operator $\hat{L}^*$ natural acts on $\bras{\gamma,j,i}\in \hil^g_{np4}$.
Consider an operator \ba
\hat{\mathscr{O}}(L):=e^{i\int d^3x \hat{T}^*\hat{h}^{sym}} \hat{L}^*e^{-i\int d^3x \hat{T}^*\hat{h}^{sym}}.\label{Diracoperator}
\ea It is obvious that $\hat{\mathscr{O}}(L)$ commutes with the scalar constraint operator $\hat{C}(N)$ in the weak sense as\ba
[\hat{\mathscr{O}}(L),\hat{C}(N)]\bras{\Psi,\gamma,j,i}=0.
\ea Also, each of such kind of operators $\hat{\mathscr{O}}(L)$ preserves
the space of solutions to the scalar constraint, since \ba
\hat{\mathscr{O}}(L)\bras{\Psi,\gamma,j,i}&=&e^{i\int d^3x \hat{T}^*\hat{h}^{sym}}\bras{\Psi_0}\otimes \hat{L}^*\bras{\gamma,j,i}\nn\\
&=&\bras{\Psi,\gamma',j',i'}.
\ea Note that the solutions \eqref{physicalsolution} are not fully diffeomorphism invariant, though they are invariant with respect to Diff($\Sigma$)$_{V_{np4}(\gamma)}$. However, as proposed in \cite{Sahlmann15}, one can average them with respect to the remaining diffeomorphisms to obtain the physical solutions to all the constraints as \ba
(\tilde{\eta})[\bras{\Psi,\gamma,j,i}]:=\sum_{[f]}U_{f}\cdot\bras{\Psi,\gamma,j,i},\label{state}
\ea where the sum ranges $[f]\in $Diff$(\Sigma)/$Diff$(\Sigma)_{V_{np4}(\gamma)}$. Since the states \eqref{state} are in the dual space of a dense subset of $\hil^{gr}_{np4}\otimes\hil^T_{np4}$, we can further introduce an inner product in their space by $<\tilde{\eta}(\Psi)|\tilde{\eta}(\Phi)>=\tilde{\eta}(\Psi)(\Phi)$. Moerover, the fully diffeomorphism invariant operators in $\hil_{gr}$ can be promoted to operators in the space of physical solutions by Eq. \eqref{Diracoperator}. Therefore, the Dirac quantization can be fully realized for this model.


To summarize, the Dirac quantization procedure of the gravity model coupled with a non-rotational dust field has been carried out in the framework of LQG without imposing any gauge fixing or solving any constraint before quantization. The main results are the following. First, the term $\mathscr{T}$ in the scalar constraint \eqref{CHC} has been regularized into the form \eqref{Thiemanntrick} suitable for the loop quantization. Second, by employing the standard loop quantization of the gravitational part and the polymer quantization of the integrated momentum of the dust field as proposed in Ref. \cite{Lewandowski15}, the scalar constraint \eqref{HamiltonGD1} has been successfully quantized. The resulting operator is well defined in the product Hilbert space $\hil^T_{np4}\otimes\hil^g_{np4}$,  which is invariant under the action of  Diff($\Sigma$)$_{V_{np4}(\gamma)}$ for any suitable graph $\gamma$. Third, the deparametrized form of the scalar constraint enables us to find a general expression of the solutions to the quantum constraint, and the observables on the space of solutions can be constructed. At last, by averaging the solutions of the scalar constraint with respect to the remaining diffeomorphisms, the Dirac quantization procedure is fully realized in LQG by this model. In comparison with the previous treatments of this model, our treatment leads to a significant difference. In the time-gauge quantization approach \cite{Husain15,Pawlowski12,Lewandowski17}, the resulting physical Hamiltonian was solely $\hat{C}^{gr}$, while in the approach of employing the diffeomorphism constraint \cite{Thiemann15,Thiemann10}, the physical Hamiltonian was $\sqrt{(\hat{C}^{gr})^2-\widehat{q^{ab}C_aC_b}}$. However, in our full Dirac quantization approach, the term $\mathscr{T}$ comes into play so that additional terms appear in the expression of the physical Hamiltonian $\hat{h}^{sym}$ by Eqs. \eqref{finalactionh2} and \eqref{finalactionh1}. In contrast to the model of coupling to scalar field \cite{Lewandowski15}, in our model the dust field dropped out of the expression of the physical Hamiltonian by regularization. This subtle change makes the construction of the solutions \eqref{physicalsolution} possible. It is desirable to further understand the low energy effective theory of the fully quantized model and derive its physical predictions. These issues are left for our future study.

\emph{Acknowledgements}--
This work is supported by NSFC with grant No. 12275087, No. 11775082, No. 11961131013, No. 11875006, No. 12275022, and ``the Fundamental Research Funds for the Central Universities''.

\appendix


\end{document}